\documentclass[aps,twocolumn,showpacs]{revtex4}
\usepackage{graphicx}
\def\pt{$p_T$ }

\def\pa{$pA$ }

\begin{document}
\title{Final-State Interaction as the Origin of the Cronin Effect}
\author{Rudolph C. Hwa$^1$ and  C.\ B.\ Yang$^{1,2}$}
\affiliation{$^1$Institute of Theoretical Science and Department of
Physics\\ University of Oregon, Eugene, OR 97403-5203, USA}
\affiliation{
$^2$Institute of Particle Physics, Hua-Zhong Normal University,
Wuhan 430079, P.\ R.\ China}
\date{\today}
\begin{abstract} Instead of adhering to the usual explanation of
the Cronin effect in terms of the broadening of the parton
transverse momentum in the initial state, we show
that the enhancement of hadron production at moderate \pt in d+Au
collisions is due to the recombination of soft and shower partons
in the final state. Such a mechanism can readily explain the
decrease of the Cronin effect with increasing rapidity.
Furthermore, the effect should be larger for protons than for
pions.

\end{abstract}
\maketitle

The conventional explanation of the Cronin effect \cite{jc}, i.e.,
the enhancement of hadron spectra at high \pt in proton-nucleus
($pA$) collisions, is that it is due to multiple scattering of
projectile  partons by the target nucleus before the production
of a minijet by a hard scattering
\cite{aa}.  All models on the effect are based on the
traditional approach to hadron production at high $p_T$, which is
to follow a hard-scattered parton by a fragmentation of that
parton.  Since in that paradigm there is nothing more one can
do with the final state, all models focus on the initial state,
and they differ only in the way the broadening of the intrinsic
traverse momentum is implemented. In this paper we
consider a drastically different approach to the problem.  We do
not assume initial-state broadening, and treat the hadronization
process in the final state by recombination \cite{dh}.  It will
be shown that the Cronin effect can be satisfactorily explained
at all centralities. In so doing we eliminate the necessity of
putting in by hand successive transverse kicks of appreciable
magnitude in the initial state.

The idea that final-state interaction may contribute to the Cronin
effect is not new \cite{tf, aai}. However, no theoretical model has
ever been proposed to demonstrate that the idea can be translated
into quantitative accounting of the effect. Here we work in the
specific framework of the recombination model and make concrete
predictions. Although the model has long been used to treat
hadronization in the fragmentation region, a number of groups have
recently found that recombination is
more important than fragmentation at small and moderate \pt $(0 <
p_T < 8$ GeV/c) at midrapidity in heavy-ion collisions at $\sqrt{s} =
200$ GeV \cite{hy}-\cite{fr}.  A central issue in the model
has always been the determination of the appropriate
distributions of partons that recombine, and it has been the
major problem to tackle for the application of the model at low
\pt in hadron-hadron
\cite{hy2} and proton-nucleus \cite{hy3} collisions.  For high \pt
we have very recently determined the distributions of shower
partons created by a hard parton \cite{hy4}, and then found that
the recombination of thermal and shower partons is crucial in
reproducing the observed spectra at moderate \pt in Au+Au
collisions
\cite{hy5}.  It is the thermal-shower recombination that also
holds the key to the Cronin effect in $pA$ collisions. It is an
effect due to the interaction between soft and hard partons during
hadronization in the final state.

Unlike the case of Au+Au collisions at RHIC where the hydrodynamical
expansion of a hot, dense system that is created leads to a large
body of thermal partons, one does not expect such a scenario in
\pa collisions. Nevertheless, there are soft partons that take the
place of the thermal partons. They can participate in the
formation of hadrons at moderate $p_T$. The number of such soft
partons decreases with increasing impact parameter and rapidity,
so our explanation of the Cronin effect naturally leads to a
reduction of the effect at lower centrality and higher rapidity,
in accordance to the observation at RHIC \cite{dd}.

The inclusive distribution for the production of pions can be
written in the recombination model, when mass effects are
negligible, in the invariant form
\cite{dh,hy2}
\begin{eqnarray} p{dN_{\pi}  \over  dp} = \int {dp_1 \over
p_1}{dp_2 \over p_2}F_{q\bar{q}} (p_1, p_2) R_{\pi}(p_1, p_2, p)
,
\label{1}
\end{eqnarray} where $F_{q\bar{q}} (p_1, p_2)$ is the joint
distribution of a $q$ and $\bar q$ at $p_1$ and $p_2$, and
$R_{\pi}(p_1, p_2, p)$ is the recombination function for forming a
pion at $p$:  $R_{\pi}(p_1, p_2, p) = (p_1p_2/p)\delta (p_1+p_2-
p)$.  Restricting $\vec{p}$ to the transverse plane, the
distribution $dN_\pi/d^2pdy|_{y=0}$, averaged over all $\phi$,
with \pt denoted by $p$, is \cite{hy5}
\begin{eqnarray} {dN_{\pi}  \over  pdp} =  {1 \over p^3} \int^p_0
dp_1F_{q\bar{q}} (p_1, p-p_1) .
\label{2}
\end{eqnarray} This equation is applicable to any of the $pp$,
$pA$ and $AB$ collision types; only $F_{q\bar{q}}$ depends on the
colliding hadron/nuclei.  In general, $F_{q\bar{q}}$ has four
contributing components represented schematically by
\begin{eqnarray} F_{q\bar{q}} = {\cal TT} + {\cal TS} + ({\cal SS})
_1 + ({\cal SS})_2 ,
\label{3}
\end{eqnarray} where $\cal{ T}$ denotes thermal distribution and
$\cal{S}$ shower distribution.  $({\cal SS})_1$ signifies two shower
partons in the same hard-parton jet, while $({\cal SS})_2$ stands for
two shower partons from two nearby jets, in a notation more self-explanatory than the one used in \cite{hy5}. For simplicity, we shall at times abbreviate $({\cal SS})_1$ by $\cal SS$.

For $pA$ collisions it may not be appropriate to refer to any
partons as thermal in the sense of a hot plasma as in heavy-ion
collisions. However, in order to maintain the same notation for
the decomposition in Eq.\ (\ref{3}) independent of the collision
types, we persist to use the symbol
$\cal{ T}$ to denote the soft parton distribution at low $k_T$,
although they  will occasionally be referred to as thermal partons,
when it is more convenient.  At low \pt
the observed pion distribution is exponential; we identify it
with the contribution of the ${\cal TT}$ term.
By writing
$\cal{ T}$ as
\begin{eqnarray} {\cal T}(p_1) = p_1{dN^{{\cal T}}_q  \over  dp_1}
= Cp_1 \exp (-p_1/T),
\label{4}
\end{eqnarray} we obtain from Eq.\ (\ref{2}) \cite{hy5}
\begin{eqnarray} {dN^{{\cal TT}}_{\pi}  \over  pdp} =  {C^2 \over
6} \exp (-p/T) \ ,
\label{5}
\end{eqnarray}
where $T$ is the inverse slope. We shall determine $C$ and $T$
by fitting the d+Au data at low $p_T$.

For shower partons we can safely ignore the term $({\cal SS})_2$ in
Eq.\ (\ref{3}) arising from two hard partons.  $\cal S$ and $({\cal
SS})_1$ involve the partons of only one shower. They are
convolutions of the hard parton distribution $f_i(k)$ with
transverse momentum
$k$ and the shower parton distributions (SPD) $S_i^j(z)$  from
hard parton
$i$ to soft parton
$j$ (and $j'$) \cite{hy4}
\begin{eqnarray} {\cal
S}(p_1)&=&\sum_i\int_{k_0}dk\,kf_i(k)\,S_i^j(p_1/k)\ ,
\label{6}\\
({\cal SS})_1(p_1,p_2)&=&\sum_i\int_{k_0}dkkf_i(k)\left\{S_i^j\left(p_1\over
k\right),S_i^{j'}\left({p_2\over k-p_1}\right)\right\}.
\label{7}
\end{eqnarray}
The integrals begin at a minimum $k_0$ below which
the pQCD derivation of $f_i(k)$ is invalid. We set $k_0=3$ GeV/c.
The curly brackets in Eq.\ (\ref{7}) signify the symmetrization of
the leading parton momentum fraction \cite{hy5}. We have assumed
in Eqs. (\ref{6}) and (\ref{7}) that the hard partons suffer no
energy losses as they traverse the cold nucleus.

The hard parton distributions $f_i(k)$ depend on the parton
distribution functions (PDF) in the proton and nucleus, and on the
hard scattering cross sections. For d+Au collisions Fries has
performed the convolution and put the inclusive distribution for
the production of a hard parton $i$ at $y=0$ and at 0-20\%
centrality in a generic form over a wide range of $k$
\begin{eqnarray}
f_i(k)\equiv {1\over \sigma_{\rm
in}}\left.{d\sigma_i^{d+Au}\over
d^2k\,dy}\right|_{y=0}=K\,A_i\left(1+{k\over k_i}\right)^{-n_i} \ ,
\label{8}
\end{eqnarray} where $\sigma_{\rm in}=40.3$ mb has been used. The
parameters $A_i, k_i,$ and $n_i$ are given in Table I \cite{rf2}.
  Nuclear shadowing effects have been taken into account through the
use of EKS98 PDF. The $K$ factor is due to higher order corrections
in pQCD. We shall set it at 2.5, as in \cite{fr,hy5,sri}.

\begin{table}

\caption{Parameters in Eq.\ (\ref{8}) \cite{rf2}.}
\begin{center}
\begin{tabular}{|l||c|c|c|c|c|c||} \hline
$i $&$u$&$d$&$s$&$\bar u$&$\bar d$&$g$\\ \hline
$A_i$&12.371&12.888&1.144&2.638&2.613&63.116\\ \hline
$k_i$&1.440&1.439&1.935&1.768&1.766&1.718\\ \hline
$n_i$&7.673&7.662&8.721&8.574&8.586&8.592\\ \hline
\end{tabular}
\end{center}
\end{table}

We calculate the three contributions $\cal TT, TS,$ and ${\cal
SS}$ to $F_{q\bar q}$ and then to $dN_\pi^{d+Au}/pdp$ in Eq.\
(\ref{2}). In the calculation there are two parameters: $C$ and $T$.
They are adjusted to fit the low \pt region of the data. The point of
view we adopt is that the soft component specified by $C$ and
$T$ is not the predictable part of our model. In treating them as free
parameters we do not compromise the predictable part of our
model, which is the magnitude of the contribution from the
$\cal TS$ component in the recombination compared to the other
components: $\cal TT$ at low \pt and ${\cal SS}$ at high $p_T$.

For d+Au collisions at RHIC, PHOBOS has published data on the \pt
spectra of charged particles, $(h^++h^-)/2$, over the range
$0.25<p_T<6.0$ GeV/c for various centralities; however, the
pseudorapidity range is $0.2<\eta<1.4$ \cite{bb}. PHENIX has
preliminary data on $\pi^+$ production over a narrower \pt range (
$<3$ GeV/c) but for $\eta=0$ \cite{phenix}. We use the latter because
they are for $\pi^+$ at $\eta=0$. Although \pt does not go above 3
GeV/c, the range is sufficient to determine $C$ and $T$, where the
$\cal TT$ contribution dominates, and where the deviation from the
exponential behavior is just enough to reveal the $\cal TS$
contribution. Our prediction is the spectra for $p_T>1$ GeV/c.

\begin{figure}[tbph]
\includegraphics[width=0.45\textwidth]{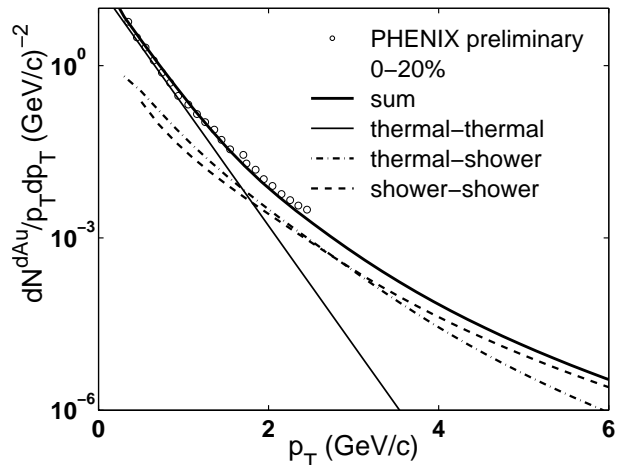}
\caption{Pion distribution in transverse momentum compared
to the data on $\pi^+$ from PHENIX \cite{phenix} on d+Au
collisions at $\sqrt s$=200 GeV and 0-20\% centrality. The three
components in the  recombination model are $\cal TT$ (light solid line), $\cal TS$ (dashed-dot line), and ${\cal SS}$ (dashed
line). Heavy solid line is the sum of all three components.}
\end{figure}

We
show in Fig.\ 1 the three components together with their sum for
0-20\% centrality. The agreement with the data  from \cite{phenix} is
good, considering the fact that only the exponential component is
adjusted to fit.  The noteworthy features of our result is that the
thermal-shower
($\cal TS$) and shower-shower (${\cal SS}$) components both become
more important than the thermal-thermal ($\cal TT$) component above
$p_T=2$ Gev/c and that $\cal TS$ is  greater than ${\cal SS}$ for
$p_T<3$ GeV/c.
The values of the parameters determined are (for 0-20\%
centrality)
\begin{eqnarray} C=12\ {\rm (GeV/c)}^{-1}, \qquad T=0.21\ {\rm
GeV/c}.
\label{10}
\end{eqnarray}

With the above value of $T$  we obtain from Eq.\
(\ref{5}) for the soft component of the pions $(p_T
\equiv p)$
\begin{eqnarray}
\left<p_T\right>&=&2T=0.42\ {\rm GeV/c},\nonumber\\
\left<p^2_T\right>&=&6T^2=0.26\ ({\rm GeV/c})^2,
\label{11}
\end{eqnarray} while Eq.\ (\ref{4}) implies for the intrinsic
transverse momentum
$k_T$ of the partons $(k_T \equiv p_1)$
\begin{eqnarray}
\left<k_T\right>&=&T=0.21\ {\rm GeV/c},\nonumber\\
\left<k^2_T\right>&=&2T^2=0.09\ ({\rm GeV/c})^2.
\label{12}
\end{eqnarray} While the numbers in Eq.\ (\ref{11}) are
conventional, the intrinsic width of the partons in Eq.\
(\ref{12}) is very small compared to what is needed in the
fragmentation models, generally $\left<k_T^2\right> > 1$
(GeV/c)$^2$, even before broadening.

The shower-shower component in Fig.\ 1 is dominant for $p_T>5$
GeV/c; it is the same as the usual contribution from parton
fragmentation
\cite{hy4,hy5}. The thermal-shower recombination is a unique
feature of our model. It makes a dominant contribution in the
$3<p_T<8$ GeV/c range in Au+Au collisions  because of the large
thermal component in the hot, dense system \cite{hy5}. Here in
the cold system only slightly excited, the values of $C$ and $T$
are lower, compared to 23.2 (GeV/c)$^{-1}$ and 0.317 GeV/c,
respectively, in
Au+Au collisions. Thus the $\cal TS$ contribution is subdued, but
still large enough not only to cause a substantial deviation of the
spectrum from exponential behavior, but also to give rise to the
Cronin effect without
large intrinsic $\left<k_T^2\right>$ broadening, as we shall
show. The ratio ${\cal TS/SS}$ is independent of the
normalization of
$f_i(k)$ and hence unaffected by the value of $K$.

\begin{figure}[tbph]
\includegraphics[width=0.45\textwidth]{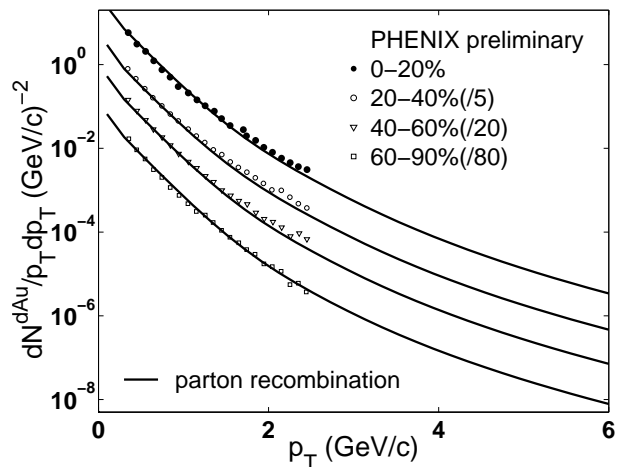}
\caption{Same as in Fig.\ 1 but for different
centralities. The inverse slope $T$ is the same in all
cases; $C$ is varied to fit the low-\pt region.}
\end{figure}

For other centralities of  d+Au collisions we fix $T$  at the value
determined for 0-20\% centrality, i.e., $T=0.21$ GeV/c,
  and use the values of $\left<N_{\rm coll}\right>$
given in
\cite{phenix} to rescale $f_i(k)$.  We adjust the value of $C$ to fit
the low \pt normalization of the data
\cite{phenix}. The results we obtain for
$dN_{\pi}^{d+Au}/pdp$  for all four centralities are shown in
Fig.\ 2 and are in good agreement with the data.  The main point
to stress is that the soft parton
$\left<k_T^2\right>$ width remains  the same at 0.09 (GeV/c)$^2$,
without being broadened by successive kicks before hard scattering.
In our approach what are different at more peripheral collisions
are the decreasing values of $C$, which are 11, 7.8, and 5.65 for
20-40\%, 40-60\% and 60-90\%, respectively. The decrease of the
density of the soft partons is a reasonable property of less
central collisions, and can be generated in a Monte Carlo code;
we merely take it from data, since it involves no new physics.
The physics issue we want to emphasize is that the
thermal-shower parton recombination is sensitive to the density
of soft partons, and that component of the hadronization product
affects the spectra in the moderately higher \pt region,
$1<p_T<4$ GeV/c.

There are inaccuracies in our calculation due to the use of the
lowest order pQCD results for the parameters in Table I, and the
SPDs that inherit the uncertainties of the FFs \cite{sk}.
   However, their effects  tend
    to cancel, if we take the ratio of the calculated spectra at
different centralities.  PHENIX has preliminary data on the
central-to-peripheral nuclear modification factor \cite{phenix}
\begin{eqnarray} R_{CP}(p_T)={\left<N_{\rm coll}\right>_{60-90\%}
dN_{\pi}^{d+Au}/p_Tdp_T(0-20\%)\over \left<N_{\rm
coll}\right>_{0-20\%} dN_{\pi}^{d+Au}/p_Tdp_T(60-90\%)}
\label{13}
\end{eqnarray}
for $p_T<6$ GeV/c.
We  can determine $R_{CP}(p_T)$
directly from the results of our calculation; it is shown in Fig.\ 3.
Evidently, the theoretical curve agrees very well with the data
\cite{phenix}. The agreement clearly lends
support to our view that the Cronin effect is due to the
recombination of soft and shower partons in the final state without
  initial-state broadening.

\begin{figure}[tbph]
\includegraphics[width=0.45\textwidth]{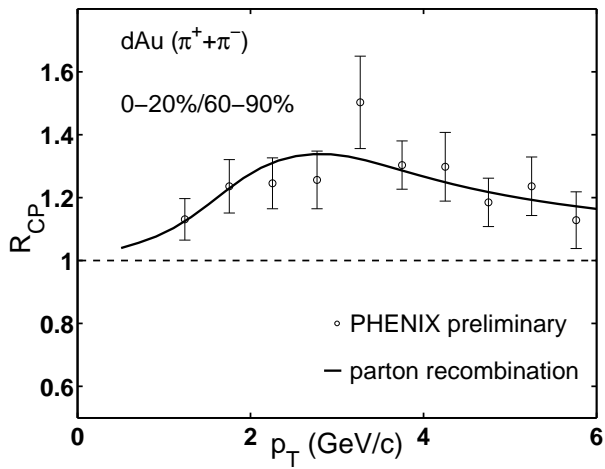}
\caption{Central-to-peripheral nuclear modification factor.
The data is from PHENIX  for d+Au collisions at $\sqrt s$=200 GeV
\cite{phenix}. The solid curve is the result of our calculation
without adjustable parameter.}
\end{figure}

The extension of our consideration to forward rapidities naturally
explains the decrease of the Cronin effect observed in d+Au
collisions \cite{rd}, since the soft parton density decreases, as one moves from the
central to the fragmentation region of the deuteron. We await the precise data in
the  low-\pt region at higher $\eta$ in order to determine the
$\eta$ dependence of $C$ in the soft component before calculating
the pion spectra.

On the basis that the thermal-shower recombination is the
component mainly responsible for the enhancement of
$R_{CP}^{\pi}$ in the intermediate range of \pt in Fig.\ 3, we
can infer qualitatively here that for proton production the
corresponding $R_{CP}^p$ should be even higher in the same
region. The reason is that for three quarks to recombine in
forming a proton, there are many more terms:
${\cal TTT+TTS+TSS+SSS}+\cdots$ \cite{hy5}, among which
the
$\cal TTS$ contribution has a quadratic dependence on $C$, and is
therefore more sensitive to centrality. Consequently, as the
collisions become more central, $dN_p^{d+Au}/p_Tdp_T$ should
receive a larger boost from
$\cal TTS$ than does  $dN_{\pi}^{d+Au}/p_Tdp_T$ from $\cal TS$,
resulting in $R_{CP}^p$  being higher than $R_{CP}^{\pi}$. The data
of PHENIX \cite{phenix} show
that such a behavior has already been observed. That behavior
is hard to interpret in a fragmentation model, since the
broadening of the parton $k_T$ width in the initial state should
be independent of what a hard parton fragments into. To determine
$R^p_{CP}$ quantitatively in our approach, we need to take the proton
mass effects into account, for which our scale invariant formalism is
inadequate. That is a task to be done elsewhere.

In conclusion we have shown, both here and in \cite{hy5}, that the
separation of final states into independent and noninteracting
soft and hard components
is invalid, except when \pt is very large. At
moderate \pt where the Cronin effect is found, the interaction
between the soft and shower partons is important. Since the density
of soft partons depends on the number of participants even in d+Au
collisions, the hadron spectra at moderate \pt depend on
centrality, when those soft partons recombine with the shower
partons.  Thus the enhancement of hadron production in more central
collisions is a final-state effect, in contrast to the usual
explanation in terms of initial-state fluctuations. The Cronin effect
may now be regarded as  another phenomenon in support of the
recombination model besides the large $p/\pi$ ratio and scaling
elliptical flow \cite{fries}. Our result also gives credence to
our assertion that shower partons form an essential component in
the  final state of a quark-gluon system produced in a heavy-ion
collision before hadronization takes place, but whose existence
has hitherto been overlooked.

       We are grateful to D.\ d'Enterria for discussions that have
stimulated our interest in this problem and for valuable
communications thereafter. We are indebted to R.\
J.\ Fries for providing us with the results of his work before
publication and for his accompanying comments that are
illuminating. We have benefitted from helpful communications with
V.\ Greco, C.\ M.\ Ko, M.\ Murray and N.\ Xu. This work was
supported, in part,  by the U.\ S.\ Department of Energy under
Grant No. DE-FG03-96ER40972 and by the Ministry of Education of
China under Grant No. 03113.


\begin{thebibliography}{99}

\bibitem{jc} J.\ W.\ Cronin {\it et.~al.}, Phys.\ Rev.\ D {\bf
11}, 3105 (1975).

\bibitem{aa} For a review see A.\ Accardi, hep-ph/0212148, and
references cited therein.

\bibitem{dh} K.\ P.\ Das and R.\ C.\ Hwa, Phys.\  Lett.\ {\bf
68B}, 459 (1977); R.\ C.\ Hwa, Phys.\ Rev.\ D {\bf 22}, 1593
(1980).

\bibitem{tf} T.\ Fields and M.\ D.\ Corcoran, Phys.\ Rev.\ Lett.\
{\bf 70}, 143 (1993).

\bibitem{aai} A.\ Airapetian et al. (HERMES
Collaboration), hep-ex/0307023.

\bibitem{hy} R.\ C.\ Hwa, and C.\ B.\ Yang, Phys.\ Rev.\ C {\bf
67}, 034902 (2003).

\bibitem{gr} V.\ Greco, C.\ M.\ Ko, and P.\ L\'{e}vai, Phys.\
Rev.\ Lett.\ {\bf 90}, 202302 (2003); Phys.\ Rev.\ C {\bf 68},
034904 (2003).

\bibitem{fr} R.\ J.\ Fries, B. M\"{u}ller, C.\ Nonaka and S.\ A.\
Bass, Phys.\ Rev.\ Lett.\ {\bf 90}, 202303 (2003);  Phys.\ Rev.\ C
{\bf 68}, 044902 (2003).

\bibitem{hy2}
       R.\ C.\ Hwa and C.\ B.\ Yang,  Phys.\ Rev.\ C {\bf 66},
025205 (2002).

\bibitem{hy3}
       R.\ C.\ Hwa and C.\ B.\ Yang, Phys.\ Rev.\ C {\bf 65},
034905 (2002).

\bibitem{hy4}
       R.\ C.\ Hwa and C.\ B.\ Yang, hep-ph/0312271.

\bibitem{hy5} R.\ C.\ Hwa and C.\ B.\ Yang, nucl-th/0401001.

\bibitem{dd} D. d'Enterria, talk given at {\it Quark Matter 2004},
17th Int'l Conf. on Ultra Relativistic Nucleus-Nucleus Collisions
(Oakland, CA, January 2004), edited by H.\ G.\ Ritter and X.\ N.\
Wang; I. Arsene et al (BRAHMS Collaboration), Phys.\ Rev.\ Lett.\
{\bf 91}, 072305 (2003).

\bibitem{rf2} R.\ J.\ Fries (private communication).


\bibitem{sri} D.\ K.\ Srivastava, C.\ Gale, and R.\ J.\ Fries,
Phys.\ Rev.\ C {\bf 67}, 034903 (2003).

\bibitem{bb} B.\ B.\ Back {\it et al.} (PHOBOS Collaboration),
Phys.\ Rev.\ Lett.\ {\bf 91}, 072302 (2003).

\bibitem{phenix} F.\ Matathias (PHENIX Collaboration), talk given
at {\it Quark Matter 2004, loc. cit}.


\bibitem{sk} S.\ Kretzer, Phys.\ Rev.\ D {\bf 62}, 054001 (2000).

\bibitem{rd} See talks by R.\ Debbe and by M.\ Murray (BRAHMS
Collaboration) at {\it Quark Matter 2004, loc. cit.}

\bibitem{fries} For a review, see R.\ J.\ Fries, {\it Quark Matter
2004, loc. cit.}

\end{thebibliography}
\end{document}